# Information Disorders, Moral Values and the Dispute of Narratives

Daniel Schwabe
dschwabe@gmail.com

## 1 Introduction

In this talk I propose a new framework to study the information disorder phenomena – mis/disinformation, conspiracy theories, gaslighting, and, to a certain extent, propaganda. To understand these various phenomena, we first look at the human communication process itself.

Among the many characterizations of human communication[1], we adopt the view that it is the information lifecycle paradigm of "production, transmission and reception of information among two or more agents". Furthermore, we argue that communication has as its ultimate goal the production or conditioning of certain actions in the receiver, which may occur in the future.

From this point of view, it is possible to characterize information disorders in terms of particular characteristics of this communication process, as proposed by Wardle&Derakhshan[2].

Our actions in life are at least partially guided by information we possess, which we acquired through various means – some through personal experience and direct observation, others received from external sources. When we need to act - e.g., forward a post, retweet, go to a rally, vote, make a donation, etc… - we take into account all the claims about relevant entities, filter out those that for some reason we don't trust, and act according to the ones we trust. In other words, the claims we believe in are treated as facts. It is important to emphasize that this process applies in the context of that specific action, and may result in different outcomes for a different action or different context.

In the current information age, we are continuously exposed to new information coming in through various channels, including TV, print media, radio, and perhaps most of all, digital media. This information can be regarded as containing claims, which are collections of statements about entities in the world. A statement basically says that a relation holds between entities, or that an entity has a certain attribute with a given value. Examples are "The US 2020 Presidential Elections were fraudulent", "Covid-19 Vaccines inoculate microchips", "Daniel Kahneman won a Nobel Prize", etc...

The incoming information we receive can be regarded as narratives that encapsulate the set of claims within it. The process of extracting such claims from the narrative – its interpretation - has been studied extensively in several disciplines – linguistics, psychology, communications, law, anthropology, etc… Given the plurality of available sources, a person will be typically exposed to several different narratives about relevant entities when s/he needs or intends to take some action.

Each narrative embodies a rationale around the claims whose goal is to make the reader accept these claims as facts. In this sense, one can say there is an ongoing dispute of narratives, because the statements about entities contained in the claims contradict each other, i.e., they cannot be true at the same time – so the reader can accept only one of them as true to carry out the action.

Fact-checking agencies strive to present alternative narratives such that the facts in the proposed narrative are more trustable. This trustworthiness may be a consequence of several non-mutually exclusive factors, ranging from the identity of the author, to surface quality of text or media, to coherence in the rationale, or to alignment with ideological/moral/ethical pre-held values.

In deciding to trust a set of claims the first consideration is whether the receiver already trusts the veracity of the statements in the claim. If so, s/he may proceed to use that information to perform the action.

However, it may be the case that the receiver either does not have any information regarding the statements, or that s/he already trusts statements that are contradictory with the statements in the claim.

In these circumstances, the receiver will resort to a variety of procedures to either accept the new set of statements as truthful, possibly substituting conflicting pre-existing statements with the newly received ones, or to reject them as being false and discarding them.

A typical approach to decide this is to inquire about the provenance of the claims - what is the source that is making them? This includes the case when the source is the recipient her/himself –the facts in the claims are the result of direct observation – with all the caveats about the fallibility of human perception.

In many situations, the user defers the truthfulness evaluation of the facts to a trusted source, so the statements contained in claims made by this trusted source are accepted as true. It should be clear that this criterion is entirely dependent on the receiver of the information, so it evidently varies from person to person.

A frequent occurrence is when a source is trusted due to social norms, e.g., for government-related actions, your "true" birthdate is the one that appears in your birth certificate, because by convention the Vital Records office has "public faith" in such matters – So any statement made by this source is true, and can be regarded as a fact.

For certain actions, however, the receiver does not trust the source, and requires additional evidence to support the veracity of the statements in order to decide whether to accept their truthfulness. This evidence is in turn a new set of claims with statements that are related to the entities in the statements of original claim; the relation must be such that the receiver will accept the original statements as true if s/he accepts the supporting statements as true. We refer to this set of supporting statements together with their source as the provenance for the original statements. At this point, the trust process is repeated with the facts in the supporting claims, and their respective provenance statements.

This recursive process creates a chain of trust that will stop when a trusted source is reached, and no further supporting statements are needed, or when resources are exhausted. We call these stopping points as *anchors* of the trust chain. In general, the same facts may require a different stopping point in this chain of trust. For example, if one wants to find the birthdate of a person to recommend a film for her/him to watch, it is sufficient to take their own statement in this respect as a fact. However, if the action involves selling alcoholic beverages, the person's

---

word cannot be trusted, and the trust chain will stop only when a document from a trusted source (i.e., an ID) is presented.

It may be the case that the trust process stops not because a trusted source has been reached, but because no additional facts can be obtained. In this case, the receiver must make a trust decision, possibly based on the trustworthiness degree associated with the source. A criterion often used is considering whether there are supporting statements in the provenance made multiple *independent* sources. Having many such sources is regarded as an indicator of higher veracity for the statements.

Summarizing, we claim that a receiver of information is continuously engaged in a cycle of receive-ingest-revise-update operations over trusted statements contained in claims. Each statement has an associated context that includes among other things its enclosing claim, the action and the provenance information.

The receiver also keeps a measure of trustworthiness associated to each source, which takes into account the action and the items (i.e., entities and relations) involved in the statement. This addresses the observation that one may trust a source on one subject (e.g., sports), but not on others (e.g., finances). Furthermore, for the same subject, the trustworthiness may vary depending on the action involved (e.g., "lending money" vs "financial advice").

A side-effect of the trust cycle of is that it contributes, in some manner, to increase or decrease the degree of trustworthiness the receiver has with respect to the sources involved in the trust chain. This degree increases as new statements contained in claims where these sources are involved become trusted. Conversely, it decreases when claims are rejected. This is consistent with the perception that trustworthiness is associated with predictability of the outcomes of an action, and/or the behavior of the agent, based on prior outcomes.

## 2 Assessing Narratives

Information disorders are ultimately characterized by the sets of trusted statements (beliefs) that are held by different groups of people, and how such statements are disseminated and adopted by people in these groups.

Confronting narratives entails comparing claims containing statements that are in contradiction with each other, and deciding which statements to believe. This decision process, according to some approaches (including the attempts at automation), can be based on seemingly rational criteria – expert consensus, evidence, and logical contradictions between the statements. But as pointed out by Vraga&Bode [3], there is a continuum on consensus both about experts (who are they, the nature of their consensus, is there any perceived bias) and about evidence (how much support they have, how concrete and how context independent they are). For many subjects, such contradictions are not easily defined or established.

Other authors have shown that this process cannot be separated from subjective, often irrational criteria. G. Lakoff postulates that the brain is hardwired to recognize metaphorical frames [4], which are tied to moral values. This has been later corroborated by experimental results in cognitive science. Furthermore, individuals can harbor conflicting values, which get applied in different contexts because they inhibit each other mutually, and only one gets activated, according to a hierarchy of values in the individual's mind. The influence of moral values on behavior is also shown in the work of D. Kahan on decisions following identity-protection and motivated reasoning [5].

In terms of the previously mentioned trust chain, following Lakoff, we posit that the anchor points in trust chains are *grounded in moral values*. In other words, the set of trusted statements are always grounded on moral values held by the individual.

Whereas information disorders were already present in traditional media, they have grown exponentially in the current digital media environment. The volume of items, and their propagation speed, compounded by automation through software agents, has urged the creation of (semi) automated tools and approaches to deal with these phenomena at scale.

The automated support for detection and characterization and countering of information disorders relies on the use of some database or Knowledge Graph that is used as a reference for facts. For the reasons already discussed, it is important that these sources include, in addition to the statements that would serve to form a competing narrative, information about the provenance of these statements, including the sources and authors.

Unfortunately, none of these sources provide any way to represent moral values, following, for example, the Moral Foundations Theory [6].

As an example of how such inclusion could help in better characterizing information disorders, consider the recently published study by the U. of Chicago CPOST [7]. One of the conclusions was that the Jan. 6 insurrectionists were driven by the perception of a great "white displacement" by blacks and hispanics in their home counties.

We show in Fig. 1 a simplified example on how this could be represented as data in a Knowledge Graph. Having such information as data may help improve both the interpretation of information being circulated and the generation of more effective competing narratives to counter information disorders.

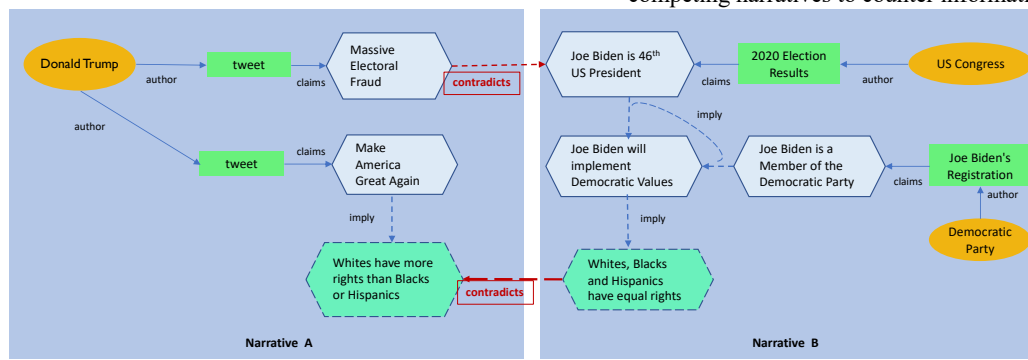

Fig. 1 Competing narratives about "Election Fraud in the 2020 US Presidential Elections"

---

[3] Vraga, E. K., & Bode, L. (2020). Defining misinformation and understanding its bounded nature: Using expertise and evidence for describing misinformation. *Political Communication*, *37*(1), 136-144

[4] Lakoff, G. (2010). *Moral politics: How liberals and conservatives think*. University of Chicago Press.

[5] Kahan, D. M. (2017). Misconceptions, misinformation, and the logic of identity-protective cognition. Yale School of Law Working Paper 164.

[6] Graham, Jesse, et al. (2013) "Moral foundations theory: The pragmatic validity of moral pluralism." *Advances in experimental social psychology*. 47, 55-130. Academic Press.

[7] https://www.nytimes.com/2021/04/06/us/politics/capitol-riot-study.html